\documentclass[useAMS,usenatbib]{mn2e}
\pdfoutput = 1

\usepackage{graphicx}
\newcommand{\kms}{\mbox{${\rm km~s^{-1}}$}}
\newcommand{\e}{\mbox{$^{-1}$}}
\newcommand{\ee}{\mbox{$^{-2}$}}
\def\simgt{\lower.5ex\hbox{$\; \buildrel > \over \sim \;$}}
\def\simlt{\lower.5ex\hbox{$\; \buildrel < \over \sim \;$}}
\def\SO{\mbox{$\sigma$\,Orionis}}

\begin{document}

\title[$850\,\mu$m survey of discs in $\sigma$\,Orionis]{A SCUBA-2 $850\,\mu$m survey of protoplanetary discs in the $\sigma$\,Orionis cluster}
\author[J. P. Williams et al.]{
Jonathan P. Williams$^1$\thanks{Email: jpw@ifa.hawaii.edu},
Lucas A. Cieza$^1$,
Sean M. Andrews$^2$,
Iain M. Coulson$^3$,
\newauthor
Amy J. Barger$^{1,4,5}$,
Caitlin M. Casey$^1$,
Chian-Chou Chen$^1$,
Lennox L. Cowie$^1$,
\newauthor
Michael Koss$^1$,
Nicholas Lee$^1$,
and David B. Sanders$^1$\\
$^{1}$Institute for Astronomy, University of Hawaii at Manoa, Honolulu, HI, 96822, USA\\
$^{2}$Harvard-Smithsonian Center for Astrophysics, 60 Garden Street, Cambridge, MA 02138, USA\\
$^{3}$Joint Astronomy Centre, 660 N. A`ohōkū Place, University Park, Hilo, HI 96720, USA\\
$^{4}$Department of Astronomy, University of Wisconsin-Madison, 475 North Charter Street, Madison, WI 53706, USA\\
$^{5}$Department of Physics and Astronomy, University of Hawaii, 2505 Correa Road, Honolulu, HI 96822, USA}

\pagerange{\pageref{firstpage}--\pageref{lastpage}} \pubyear{2013}
\maketitle

\label{firstpage}

\begin{abstract}
We present the results from a large $850\,\mu$m survey of the
\SO\ cluster using the SCUBA-2 camera on the
James Clerk Maxwell Telescope.  The $0.5^\circ$ diameter circular
region we surveyed contains 297 young stellar objects with an age estimated
at about 3\,Myr.  We detect 9 of these objects, 8 of which have
infrared excesses from an inner disc.
We also serendipitously detect 3 non-stellar sources at $>5\sigma$
that are likely background submillimetre galaxies.
The 9 detected stars have inferred disc masses ranging from
5 to about $17\,M_{\rm Jup}$, assuming similar dust properties
as Taurus discs and an ISM gas-to-dust ratio of 100.  There is a net
positive signal toward the positions of the individually undetected
infrared excess sources indicating a mean disc mass of $0.5\,M_{\rm Jup}$.
Stacking the emission toward those stars without infrared excesses
constrains their mean disc mass
to less than $0.3\,M_{\rm Jup}$, or an equivalent Earth mass in dust.
The submillimetre luminosity distribution is significantly different
from that in the younger Taurus region, indicating disc mass evolution
as star forming regions age and the infrared excess fraction decreases.
Submillimeter Array observations reveal CO emission toward 4 sources
demonstrating that some, but probably not much,
molecular gas remains in these relatively evolved discs.
These observations provide new constraints on the dust and gas mass of
protoplanetary discs during the giant planet building phase and provide a
reference level for future studies of disc evolution.
\end{abstract}


\section{Introduction}
The ever growing number of exoplanet discoveries and their detection
statistics illustrate that the natural endpoint of a circumstellar disc
is a planetary system \citep{2013Sci...340..572H}.
Infrared observations of young stellar objects (YSOs) are highly sensitive to
small amounts of dust and numerous surveys have now established that
discs are almost ubiquitous at early times and have a median lifetime
of a few Myr \citep{2009AIPC.1158....3M}.
This timescale provides a critical constraint on planet formation
theories \citep{2005A&A...434..343A},
but it is equally important to measure the dust and gas content of
the discs in order to understand the diversity of exoplanetary systems.

Millimeter wavelength observations are the best way to measure disc
dust masses as the continuum emission is optically thin and only
mildly dependent on temperature.
Gas masses are harder to determine but the bulk of the disc mass
is cold and molecular and is therefore best traced by rotational lines
in the same wavelength range.  The emission is weak at these long
wavelengths, however, and the number of discs that have been detected
is relatively small compared to infrared surveys.
The situation is particularly acute for discs
in regions that are older than $\sim 1$\,Myr, which severely limits our
understanding of disc evolution \citep{2011ARA&A..49...67W}.

The SCUBA-2 camera on the 15\,m James Clerk Maxwell Telescope (JCMT)
provides a unique combination of high sensitivity and large field of
view at submillimetre ($450\,\mu$m and $850\,\mu$m) wavelengths
that makes possible, for the first time,
unbiased surveys of young stellar clusters for faint emission from
protoplanetary discs on a comparable scale to Spitzer Space Telescope
surveys \citep{2013MNRAS.430.2513H}.
Here, we present the observations and analysis of a SCUBA-2 survey
of the \SO\ cluster at $850\,\mu$m.

The cluster is named after its brightest member, an O9V star in a
quadruple system and a member of the Orion OB1b association.
\citet{2008hsf1.book..732W} provide a detailed summary of its properties in
the Handbook of Star Forming Regions.
The salient features of the region for this work are as follows:
it consists of several hundred YSOs from substellar to over $20\,M_\odot$,
with a distribution that is consistent with the Salpeter IMF,
amounting to a total stellar mass of about $225\,M_\odot$
\citep{2004AJ....128.2316S}.
There is little nebulosity around the cluster and the reddening along
the line of sight is small, E(B-V)$\sim 0.05-0.08$\,mag
\citep{2008AJ....135.1616S}
both suggesting that the natal molecular cloud has almost entirely dispersed.

The cluster is too distant for a reliable Hipparcos parallax
and, as protostellar luminosity depends both on age
and distance, both these quantities are uncertain and
inversely correlated.  In this work, we adopt the values from
\citet{2008hsf1.book..732W}
for the cluster age, $\sim 3$\,Myr, and distance 420\,pc.
Based on the tightness of the cluster locus on a colour-magnitude
diagram, the YSO age spread is fairly small, $\sim 1-2$\,Myr.

Although \SO\ is considerably more distant than the younger,
well studied Taurus and $\rho$\,Ophiuchus star forming regions,
it is one of the closest clusters in an intermediate evolutionary
phase when discs are dispersing.
Further, as a member of the Orion complex,
it comprises part of the closest {\it massive} star forming region.
Indeed, for a range of reasons from the distribution of
stellar associations to cosmochemistry, Orion is considered a
better analog of the Solar birthplace
\citep{2010ARA&A..48...47A, 2012A&A...545A...4G}.
The relatively high stellar density partly mitigates the greater distance
as most of the cluster sources lie within $\sim 0.5^\circ$ (3.5\,pc)
of each other and can be observed within a single SCUBA-2 field.

We use the cluster membership as defined by
\citet[hereafter H07]{2007ApJ...662.1067H}
who surveyed the region with Spitzer to identify point sources
with characteristic optical-infrared colour of protostars.
This list of 336 sources is complete to about $0.15\,M_\odot$.
The fraction of stars with optically thick discs
is $26.6\pm 2.8$\%, which is consistent for its estimated $\sim 3$\,Myr
age with the well characterised infrared disc lifetime plot
\citep{2009AIPC.1158....3M}.
There is no significant difference in the spatial distribution
of bright, higher mass stars or disc bearing stars relative to
fainter, lower mass stars and discless stars.  Thus our SCUBA-2
map provides an essentially unbiased measure of the dust mass
for a large sample of discs as they evolve from primordial to
debris through the process of planet formation.

The SCUBA-2 observations and followup interferometry are described in
\S\ref{sec:obs}.  The results are presented in \S\ref{sec:results}.
Although only a handful of discs are detected, several of these
exhibit molecular line emission which suggests that the evolution of
the dust and gas content are broadly correlated.
We compare our data with millimetre surveys of other regions
and discuss the implications for disc evolution and planet
formation timescales in \S\ref{sec:discussion}.
Finally, we summarise the paper in \S\ref{sec:summary}.

\section{Observations}
\label{sec:obs}
\subsection{SCUBA-2}
We observed a circular region with diameter $0.5^\circ$
(about the diameter of the full Moon)
toward the \SO\ cluster at $850\,\mu$m.
The data were taken in queue mode over numerous
observing runs from October 2011 to January 2013
(program IDs: M11BH02A, M12AH02A, M12BH47A; see Table~\ref{tab:obslog})
in median (JCMT band 3) weather conditions, defined by the zenith
optical depths at 225\,GHz lying between 0.08 and 0.12.  This corresponds
to precipitable water vapor levels $\sim 2-3$\,mm
and zenith optical depths at $850\,\mu$m $\sim 0.25-0.35$.
The total on-source integration time was 31 hours.

The pointing center, RA=05:39:00.0, Dec=-02:33:00.0,
lies a couple of arcminutes northeast of the brightest cluster member,
\SO, and was chosen to maximize the number of infrared
identified YSOs in the map.  The {\it pong-1800} mapping mode was used to
provide nearly uniform sensitivity over the large circular field.
The data reduction was carried out at the Joint Astronomy Center
using the Dynamic Iterative Map Maker in the STARLINK/SMURF
software package \citep{2013MNRAS.430.2545C}.
As there was expected to be little extended $850\,\mu$m
emission and because we were most interested in discs around YSOs,
which would be point sources at the $15''$ SCUBA-2 resolution,
we used the blank field configuration
for the map maker that is most commonly used for deep field
extragalactic surveys.

\begin{figure}
\centering
\includegraphics[width=70mm]{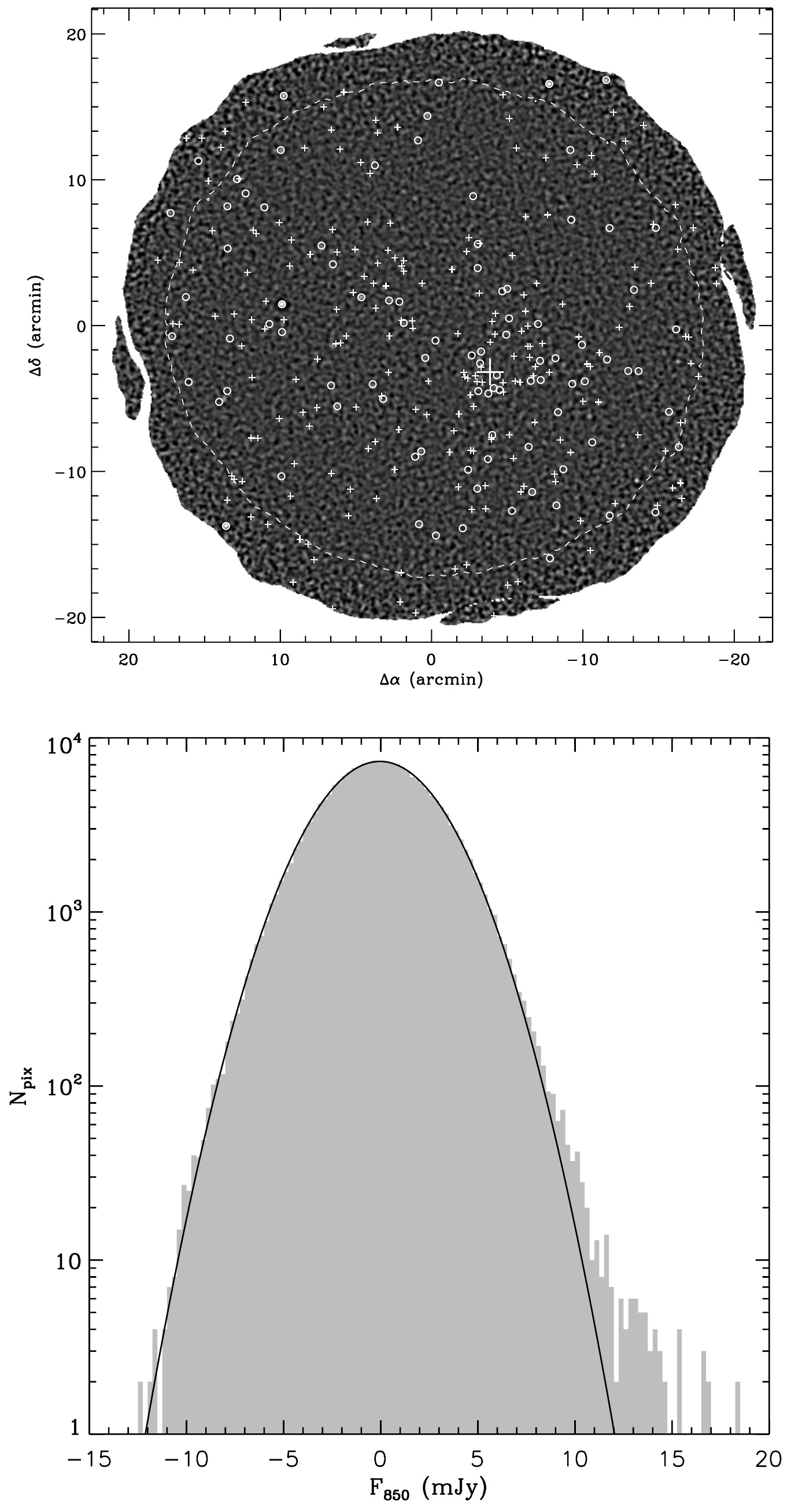}
\caption{The upper panel is the SCUBA-2 $850\,\mu$m map with
the locations of the 297 cluster members within the surveyed
region (circles for Class I, II, and TD;
crosses for Class III, EV, and DD).  The large cross indicates
the location of the O9 star, \SO. The lower panel is the histogram
of flux densities within the dashed contour of the SCUBA-2 map.
The gaussian fit has a dispersion of 2.9\,mJy\,beam\e.
The tail at high flux densities shows significant detections.}
\label{fig:scubahist}
\end{figure}

Each night was reduced individually and the maps were then co-added with
weights proportional to the inverse square of the map noise.
The flux density scale was determined through regular observations of bright
point sources, most commonly Uranus, Mars, and protoplanetary nebulae
CRL\,618 and CRL\,2688.  Following the recommended procedure in online
documentation\footnote[1]{http://www.jach.hawaii.edu/JCMT/continuum/scuba2/\\
scuba2\_relations.html}, 
we reduced these sources with the
same (blank field) map maker configuration file to derive the
flux calibration factor (FCF).  These produced a range of values,
FCF $= 715\pm 70$\,Jy\,pW\e\,beam\e.
with no clear trend on observation date, weather, or elevation.
We therefore simply multipled the co-added map by the mean value,
715\,Jy\,pW\e\,beam\e, and consider the derived flux densities to
have a calibration uncertainty of 10\%.
This FCF is about 30\% higher than the value derived for observations
of bright or extended sources by \citet{2013MNRAS.430.2534D}
due to the additional filtering that is carried out for blank fields.
Finally, we applied a matched beam filter
to smooth the data and maximize the sensitivity to point sources.
Figure~\ref{fig:scubahist} shows the final map and a histogram of
flux densities.  The rms of the central region of the map, covering a
total of 938 square arcminutes is 2.9\,mJy\,beam\e.
The noise increases toward the edges but has a usable area, within
which we could make reliable detections, of 1329 square arcminutes
and contains 297 known cluster members.

The weather conditions were too poor for sensitive $450\,\mu$m
observations and we do not use the data from those subarrays
in our analysis.

\subsection{Submillimeter Array}
A handful of detections were apparent once the first set of SCUBA-2
data were analyzed.  We followed these up with interferometric
observations on the Submillimeter Array (SMA) in the $1300\,\mu$m band.
These provided a confirmation of the SCUBA-2 detections and extended the
SED to slightly longer wavelengths.  Furthermore, the correlator was
configured to observe the 230\,GHz $J=2-1$ transition of CO
(and its isotopologues) so we could also search for molecular gas emission.

The first set of observations were short ``filler'' tracks that were
carried out between queue scheduled programs.  The second set were
dedicated tracks (program 2012A-H002) to observe four sources together,
observed in excellent weather conditions, and produced
much better visibility plane coverage and sensitivity.
The dates, observed sources, and atmospheric opacity are listed in
Table~\ref{tab:obslog}.  All observations
were carried out in the compact configuration (20--70 meter baselines)
providing an angular resolution $\sim 2-3''$.  The observational setup
was typical for the SMA, consisting of a long
integration on a passband calibrator (3C84 or 3C279) and interleaving
20 minute integrations on the targets with $2-3$\, minute integrations
on one or two gain calibrators (J0607-085, J0501-019).  Flux calibration
was carried out through observations of bright solar system targets
(Callisto or Uranus).
The data reduction also followed standard practice, using the MIR software
package to calibrate the visibilities and MIRIAD for final imaging.

\begin{table}
\caption{Observing log}
\label{tab:obslog}
\begin{tabular}{@{}cccc@{}}
\hline
Date  &  Source  &  $t_{\rm int}^{\rm a}$  &  $\tau_{225}$\\
\hline
\multicolumn{4}{c}{JCMT} \\
\hline
20111025  &  $\sigma$\,Ori  &  3.8  & 0.10  \\
20111109  &  $\sigma$\,Ori  &  1.0  & 0.13  \\
20111110  &  $\sigma$\,Ori  &  2.1  & 0.12  \\
20111218  &  $\sigma$\,Ori  &  0.5  & 0.11  \\
20111227  &  $\sigma$\,Ori  &  2.4  & 0.08  \\
20120121  &  $\sigma$\,Ori  &  2.8  & 0.10  \\
20120203  &  $\sigma$\,Ori  &  1.5  & 0.08  \\
20121224  &  $\sigma$\,Ori  &  3.1  & 0.11  \\
20130105  &  $\sigma$\,Ori  &  3.4  & 0.11  \\
20130106  &  $\sigma$\,Ori  &  3.0  & 0.11  \\
20130107  &  $\sigma$\,Ori  &  3.5  & 0.09  \\
20130108  &  $\sigma$\,Ori  &  4.5  & 0.09  \\
\hline
\multicolumn{4}{c}{SMA} \\
\hline
20111204  & V510\,Ori                                  &  2.0   & 0.15  \\
20120324  & J0539-0230                                 &  0.7   & 0.12  \\
20121112  & HD294268, V2731\,Ori, V602\,Ori, V606\,Ori &  5.2   & 0.07  \\
20121114  & HD294268, V2731\,Ori, V602\,Ori, V606\,Ori &  4.0   & 0.05  \\
20121115  & HD294268, V2731\,Ori, V602\,Ori, V606\,Ori &  2.0   & 0.05  \\
\hline
\multicolumn{4}{l}{$^{\rm a}$On-source integration time in hours.}
\end{tabular}
\end{table}

\section{Results}
\label{sec:results}
\subsection{Continuum emission}
\subsubsection{SCUBA-2}
\label{sec:results:s2}
297 of the 336 total cluster members identified by H07
lie within the SCUBA-2 map boundary.
The $850\,\mu$m flux densities were determined from a
beam-weighted average within a $15''$ diameter circular region centered
on their 2MASS positions.  The local noise level was measured from the
rms within the annulus between $20''$ to $60''$ in radius.
In general the cluster members are sufficiently separated that there
is no confusion in these measurements.
A few sources lie within $1'$ of each other which could produce a slight
increase in their noise estimates.  Only one source, ID 1005, lies this
close to a detected source (J0539-0230), and the local noise was therefore
measured manually.
The results are tabulated in Table~\ref{tab:allfluxes}.

\begin{figure*}
\vbox to 160mm{\vfil
\includegraphics[width=160mm,angle=90]{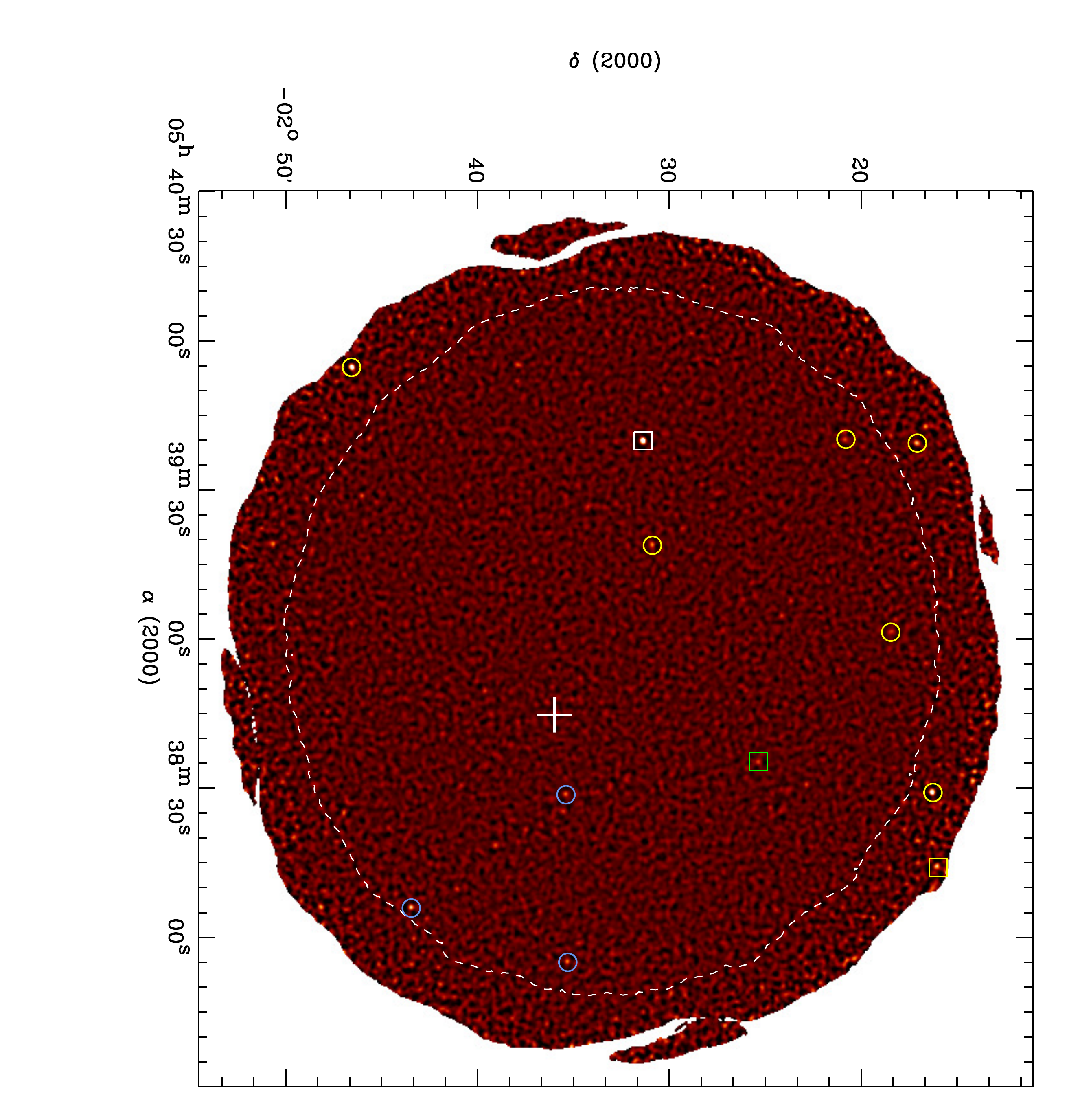}
\caption{A larger view of the SCUBA-2 $850\,\mu$m map with an overlay
showing the detections of one Class I source (white square),
6 Class II sources (yellow circles), one transition disc (yellow square),
one Class III YSO (green square), and three possible extragalactic
sources (blue circles).
The white cross indicates the location of \SO.
The dashed contour outlines the region within which
the noise level is an approximately uniform 2.9\,mJy\,beam\e.  The noise
level increases beyond this to about 4.4\,mJy\,beam\e\ at the map boundary.}
\vfil}
\label{fig:scubaplot}
\end{figure*}

\begin{figure}
\centering
\includegraphics[width=80mm]{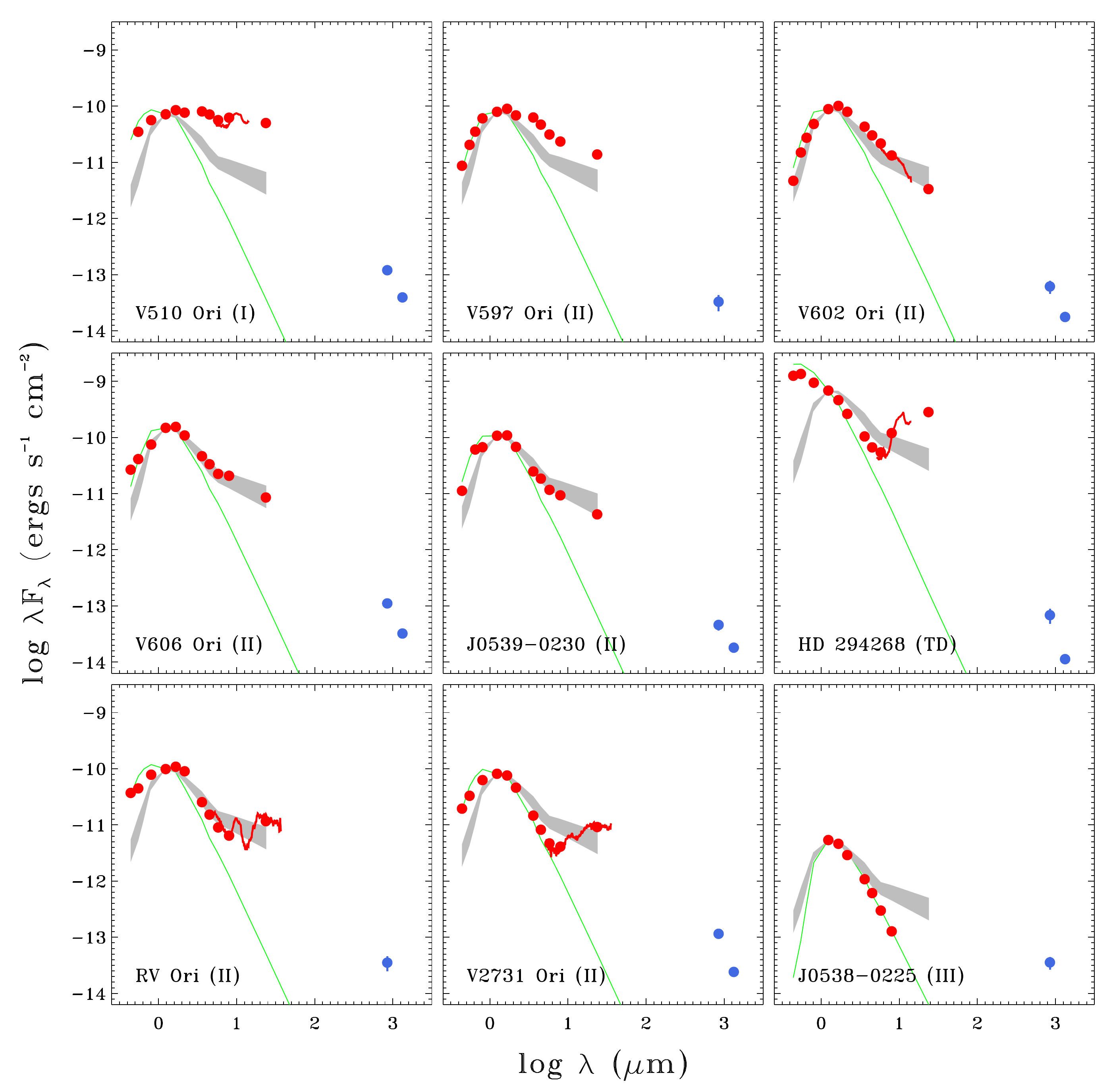}
\caption{Infrared-millimetre wavelength spectral energy distributions
of the nine detected YSOs in the SCUBA-2 map.
Error bars are shown when they are larger than the symbol size.
The optical photometry is from \citet{2004AJ....128.2316S},
the near-infrared from the 2MASS survey, and the mid-infrared
from Spitzer observations by H07.  Where available, the Spitzer
IRS spectrum is plotted with a red line.
The SCUBA-2 and SMA (where available)
flux densities are plotted as blue circles.
The green lines are the
model stellar photospheres based on spectral types where known
else optical colour, normalised to the J band ($1.1\,\mu$m).
The grey region outlines the similarly scaled lower and upper quartile
SED for optically thick discs in the cluster.}
\label{fig:sed}
\end{figure}

9 sources were detected at $>3\sigma$ significance.  They are listed in
Table~\ref{tab:discfluxes} and plotted in Figure~\ref{fig:scubaplot}.
All but one of the detections have strong infrared excesses.
H07 characterised the YSO evolutionary states by the
mid-infrared SED slope across the IRAC bands, $3.6-8.0\,\mu$m,
a shorter wavelength range than is typical.
The breakdown of detections in each classified group is as follows:
1 out of 2 Class I;
6 out of 83 Class II;
1 out of 6 transition discs (TD);
0 out of 13 evolved discs (EV);
1 out of 191 Class III;
and 0 out of 2 debris discs (DD).
Due to the inherent fuzziness of the single parameter SED slope
categorization, we follow H07 and group Class I, II, and TD together
as optically thick discs from here on.
The SEDs of the detected YSOs are shown in Figure~\ref{fig:sed}
with photospheres that correspond to the spectral types in
Table~\ref{tab:discfluxes} or, where unknown, that best match the optical
colour. IRS spectra from the Spitzer Archive are plotted when available.
For reference, the lower to upper quartile range of the infrared SED
for optically thick discs
(from Table 4 of H07, scaled to the J band for each source)
is also over-plotted.
There is a wide range of SED types, from the flat spectrum source
V510\,Ori (labeled Class I by H07),
to the apparently discless Class III source J0538-0225.
Three sources have a pronounced mid-infrared dip with little or
no excess at short wavelengths, and strong emission at $24\,\mu$m.
H07 labels only one of these, HD\,294268, a transition disc.

With the exception of the Class III source, J0538-0225,
the SCUBA-2 detections are optically brighter than the median.
This is not surprising given that more massive stars tend to
have more massive discs, at least initially \citep{2013ApJ...771..129A}.
We do not find a similar demarcation between detected and undetected
discs in terms of infrared colour, however.
The diversity of SED types suggests that there is no detailed
correlation between the inner disc geometry and bulk dust
mass, as noted before by \citet{2008ApJ...686L.115C}
and attributed to a range of evolutionary pathways as the
mass in small grains diminishes.

\begin{table}
\caption{SCUBA-2 flux densities$^{\rm a}$}
\label{tab:allfluxes}
\begin{tabular}{@{}lccccc@{}}
\hline
ID$^{\rm b}$ & Class$^{\rm b}$ & R.A. (J2000) & Dec. (J2000) & $F_{850}$ & $\sigma_{850}$ \\
             &                 &     (deg)    &      (deg)   &   (mJy)   &      (mJy)     \\
\hline
 164  &  III  &  84.43706  &  -2.49920  &   3.57  &  4.76  \\
 165  &  III  &  84.43856  &  -2.48111  &  -7.17  &  4.48  \\
 194  &  III  &  84.45671  &  -2.60506  &  -0.89  &  3.06  \\
 209  &  III  &  84.46288  &  -2.43542  &  -2.17  &  3.76  \\
 214  &  III  &  84.46497  &  -2.59045  &   3.95  &  3.55  \\
\hline
\multicolumn{6}{l}{$^{\rm a}$Only the first few rows of this table are shown, the rest}\\
\multicolumn{6}{l}{will be made available electronically.}\\
\multicolumn{6}{l}{$^{\rm b}$Source identification and class from H07.}
\end{tabular}
\end{table}

\begin{table*}
\centering
\begin{minipage}{140mm}
\caption{Detected discs}
\label{tab:discfluxes}
\begin{tabular}{@{}lcccccrrr@{}}
\hline
Name & ID$^{\rm a}$ & SpT & Class$^{\rm a}$ & R.A. (J2000) & Dec. (J2000) & $F_{850}$ & $F_{1300}$ & $F_{\rm CO2-1}$  \\
     &              &     &                 &     (deg)    &     (deg)    &    (mJy)  &    (mJy)   & (Jy km s$^{-1}$) \\
\hline
 V510 Ori    &  1153  & \ldots  &   I  &  84.91600  & $-2.52280$  &  $33.9 \pm 3.3$  &  $17.4 \pm 1.2$  &  $ 0.56 \pm 0.16$  \\
 V2731 Ori   &   540  & \ldots  &  II  &  84.62143  & $-2.27104$  &  $32.6 \pm 4.8$  &  $10.7 \pm 0.8$  &  $ 0.95 \pm 0.13$  \\
 V606 Ori    &  1274  & \ldots  &  II  &  84.97777  & $-2.77618$  &  $31.4 \pm 4.8$  &  $14.3 \pm 0.8$  &  $ 0.36 \pm 0.07$  \\
 HD 294268   &   411  & F8      &  TD  &  84.55875  & $-2.26657$  &  $19.4 \pm 5.4$  &  $ 5.0 \pm 0.8$  &  $< 0.30$          \\
 V602 Ori    &  1152  & M3      &  II  &  84.91411  & $-2.28464$  &  $17.4 \pm 4.2$  &  $ 7.8 \pm 0.8$  &  $ 0.34 \pm 0.08$  \\
 J0539-0230  &   984  & K9      &  II  &  84.82854  & $-2.51480$  &  $13.0 \pm 2.7$  &  $ 8.0 \pm 1.0$  &  $< 0.81$          \\
 J0538-0225  &   609  & M6      & III  &  84.64726  & $-2.42280$  &  $10.1 \pm 2.5$  &  \ldots          &  \ldots            \\
 RV Ori      &  1156  & \ldots  &  II  &  84.91740  & $-2.34667$  &  $10.0 \pm 2.9$  &  \ldots          &  \ldots            \\
 V597 Ori    &   844  & \ldots  &  II  &  84.75575  & $-2.30771$  &  $ 9.3 \pm 3.1$  &  \ldots          &  \ldots            \\
\hline
\multicolumn{9}{l}{$^{\rm a}$ Source identification number and class from H07.}
\end{tabular}
\end{minipage}
\end{table*}

\begin{figure}
\centering
\includegraphics[width=80mm,angle=90]{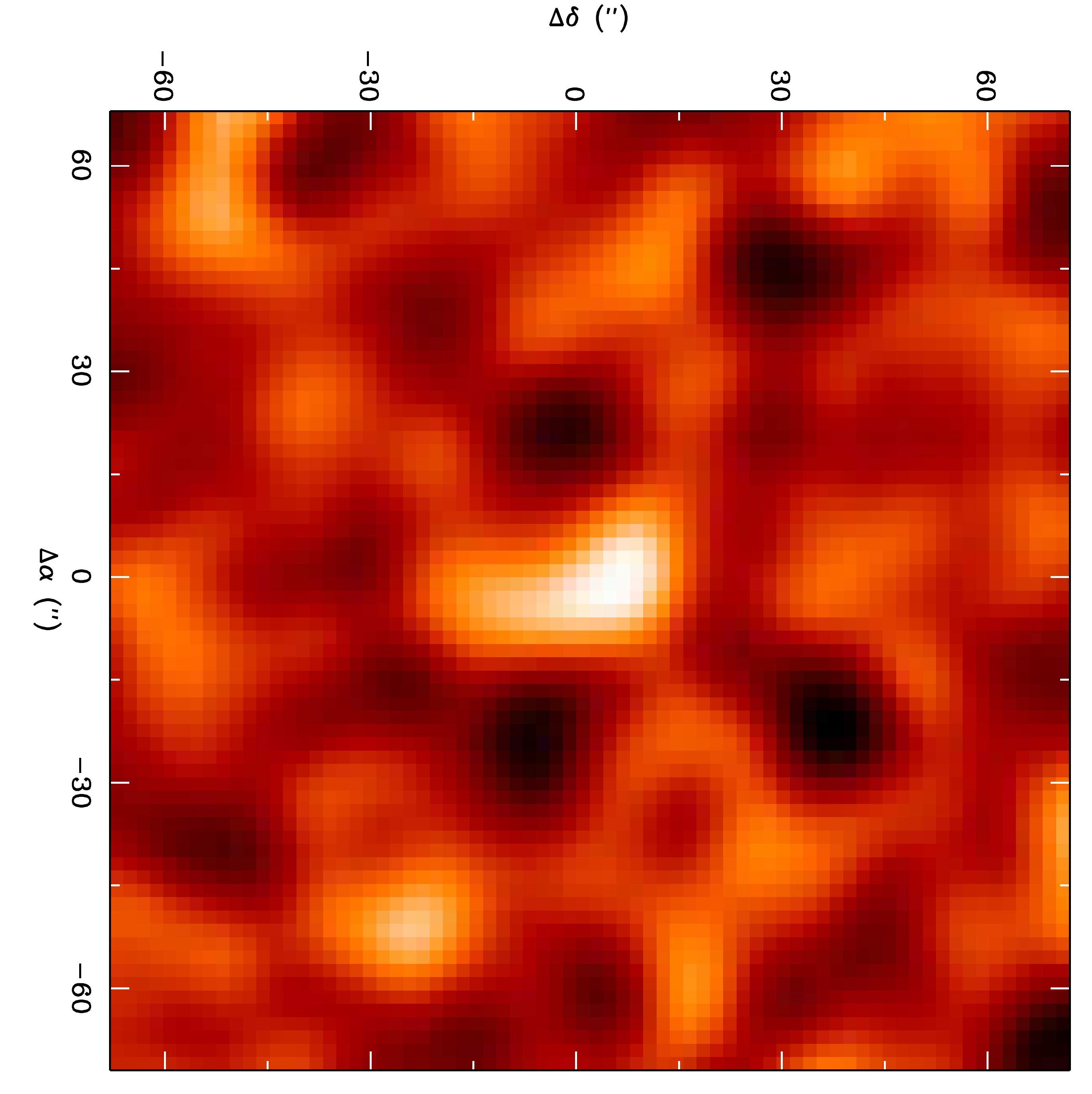}
\caption{Stacked map of SCUBA-2 $850\,\mu$m emission toward
the 83 Class II YSOs that were not detected individually.
The central peak flux density is 1.3\,mJy\,beam\e,
significant at the $4\sigma$ level,
and confirms the analysis of the photometry statistics.}
\label{fig:stack}
\end{figure}

\subsubsection{Stacking analysis}
Most of the cluster members, including those with infrared excesses
from optically thick discs, are not detected in the SCUBA-2 map.
However, by analyzing the statistics of the measured flux densities
toward their positions in the map (as listed in Table~\ref{tab:allfluxes}),
we can constrain average properties of the ensemble.
There are 288 sources that were not individually
detected at $3\sigma$ significance.
Based on the H07 classification,
83 of these are optically thick discs
(mostly Class II but including one Class I and 5 TD)
and 190 are Class III,
the remainder being 13 optically thin ``evolved discs''
(also termed anemic discs by \citet{2006AJ....131.1574L}) and 2 debris discs.
The mean flux density and standard error of the mean for the two main subgroups
are listed in Table~\ref{tab:stack}.
There is a positive mean flux density toward the
optically thick discs, significant at the $3.3\sigma$ level.
We directly verified this result by shifting the map to each source
location and adding to produce the stacked map in Figure~\ref{fig:stack}.
The peak flux density of this map is 1.3\,mJy\,beam\e, which is
slightly higher than (but consistent with) the statistical mean
of the photometry.

There is no significant detection in the stacked map or
in the statistics of the photometry of the Class III sources.
As these lack an infrared excess, it may not be surprising that we
do not detect these sources, but these data provide
a stringent limit to their average $850\,\mu$m flux density
and therefore dust mass (see \S\ref{sec:discussion}).
These results on the collective properties of the discs provide a useful
reference point for evolutionary studies in this relatively evolved cluster
where 74\% of discs have dispersed.

\begin{table}
\caption{Stacking analysis}
\label{tab:stack}
\begin{tabular}{@{}cccc@{}}
\hline
     Type         & $N$  &  $\mu$  & $\sigma_\mu$ \\
                  &      &  (mJy)  &    (mJy)     \\
\hline
 Optically thick  &  83  &   1.03  &  0.30  \\
 Class III        & 190  &   0.10  &  0.15  \\
\hline
\end{tabular}
\end{table}

\subsubsection{Submillimeter Array}
Six of the SCUBA-2 sources were observed with the SMA at
$1300\,\mu$m and all were clearly detected
(column 8 of Table~\ref{tab:discfluxes}).
The resolution of these data, $\sim 2.5'' = 1000$\,AU, was too
coarse to resolve the discs.  However, it verifies the continuum
detection and allows us to measure the SED slope,
$\alpha_{\rm mm}$, where $F_\nu\propto\nu^\alpha_{\rm mm}$.
The six discs have similar, very shallow slopes with mean
value $\langle\alpha_{\rm mm}\rangle=1.9\pm 0.5$.
This is close to the blackbody value of 2 which suggests
significant grain growth, at least several millimetres in size
\citep{2006ApJ...636.1114D}.
To go beyond suggestion, however, requires longer wavelength
observations \citep[e.g.,][]{2010A&A...512A..15R} and,
ideally, resolved multi-wavelength data \citep{2012ApJ...760L..17P}.

The SMA target list was defined from the early SCUBA-2 maps
and therefore biased toward what turned out to be the
strongest sources.  Unfortunately, we are not yet able to
independently verify the three faintest $850\,\mu$m sources.
These each lie within the central region of the map
where the noise variations are small but, nevertheless,
there is a non-negligible probability of a chance
alignment of a $3\sigma$ noise spike with the location
of a cluster member.
In particular, there is about a 25\% chance that a
false detection would be found toward one of the
191 Class III sources.  Whereas the detection of J0538-0225
is potentially very interesting, it requires independent confirmation.

\subsection{CO emission}
\begin{figure}
\centering
\includegraphics[height=150mm]{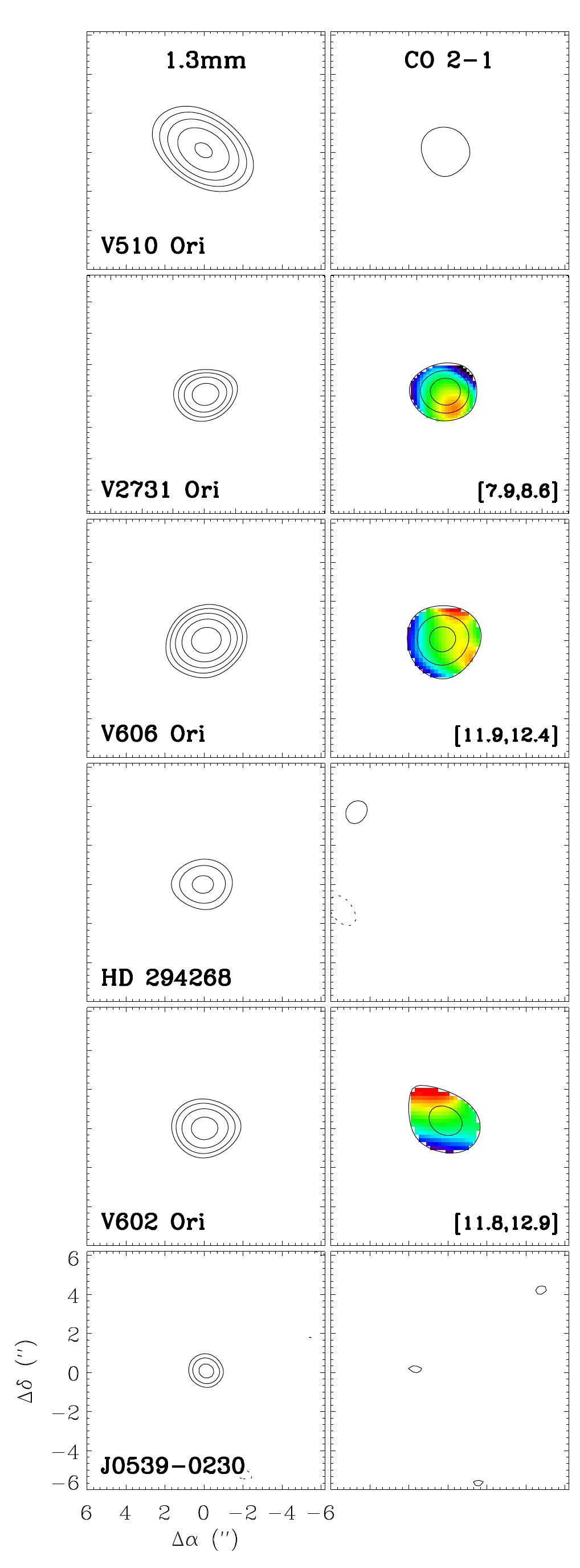}
\caption{$1300\,\mu$m continuum (left panel) and CO 2--1 line emission
toward the 6 discs observed with the SMA.  For each plot the
contour levels begin at three times the rms (see Table~\ref{tab:discfluxes})
and negative emission is shown as dotted contours.
The continuum contours increase by a multiplicative factor of 1.5
for each step.
The line contours are linear with steps of twice the rms.
For three sources with sufficiently strong line emission,
the first moment map is underlaid in colour for the velocity
range indicated in square brackets.}
\label{fig:sma}
\end{figure}

We detected CO 2--1 emission toward 4 of the 6 SMA sources
(column 9 of Table~\ref{tab:discfluxes}).
In three cases, the emission was strong enough to make a
first moment map and reveal a velocity gradient across the
(unresolved) disc.  Figure~\ref{fig:sma} shows the $1300\,\mu$m continuum
and CO maps.  There is no clear correlation between the
presence of detectable CO and other properties of the sources
although the sample is obviously very small.
The CO emission is almost surely optically thick and its strength
depends more on the disc temperature than the gas mass.
However, the simple fact of its detection tells us that some
molecular gas remains in these relatively old discs.
The lower sideband of the SMA included the same $J=2-1$ transition
of the isotopologues, $^{13}$CO and C$^{18}$O, which could
provide a more accurate measure of the disc gas mass,
but these were not detected in any source.

\subsection{Background sources}
With a depth of 2.9\,mJy\,beam\e\ over an area of
938 square acrminutes (within the dashed contour of
Figure~\ref{fig:scubaplot}), our SCUBA-2 map of the
\SO\ cluster can also be considered a
moderately deep large field to search for submillimetre galaxies (SMGs).
First detected with the original SCUBA camera
\citep[e.g.,][]{1997ApJ...490L...5S, 1998Natur.394..248B, 1998Natur.394..241H},
and also seen in the new SCUBA-2 surveys
\citep{2013ApJ...762...81C, 2013arXiv1302.2619C, 2013MNRAS.432...53G},
these prodigious star forming, dusty galaxies contribute significantly
to the star formation rate density at high redshifts and are likely a
key evolutionary phase in the buildup of massive galaxies.

There are about 13000 independent elements in our map
so a blind $3\sigma$ cutoff will have many false
positives and we even expect one $4\sigma$ noise spike.
We therefore conservatively choose a $5\sigma$
threshold away from known cluster members to search for
background sources.  Three such objects are identified,
shown by the blue circles in Figure~\ref{fig:scubaplot}
and tabulated in Table~\ref{tab:smg}.
We have not found any counterparts within $15''$ of these sources
at any other wavelength and consider them to be SMGs.

The latest SCUBA-2 $850\,\mu$m number counts model from
Chen et al. 2013 (submitted) predict $3.3^{+5.7}_{-2.2}$
sources greater than 15\,mJy within the area of our map
which is entirely consistent with our observations.

\begin{table}
\caption{Submillimetre Galaxies}
\label{tab:smg}
\begin{tabular}{@{}lccc@{}}
\hline
Name & R.A. (J2000) & Dec. (J2000) & $F_{\rm 850}$ \\
     &     (deg)    &     (deg)    &     (mJy)     \\
\hline
 SMG\,J053805.9-024327  &  84.52458  &  $-2.72428$  &  $24.9 \pm 3.6$ \\
 SMG\,J053755.0-023518  &  84.47917  &  $-2.58825$  &  $17.0 \pm 3.2$ \\
 SMG\,J503828.7-023523  &  84.61958  &  $-2.58986$  &  $15.8 \pm 2.8$ \\
\hline
\end{tabular}
\end{table}

\section{Discussion}
\label{sec:discussion}
The wide-field mapping capability of SCUBA-2 has allowed us to make
sensitive $850\,\mu$m measurements toward almost 300 members of the
\SO\ cluster.  We were able to follow up the brightest
sources with SMA interferometry and search for gas.
Here we convert the measured flux densities to physical quantities and
compare with other surveys to learn about disc evolution.

\subsection{Disc masses}
With limited information about the stellar luminosity and temperature,
and unresolved disc images, we cannot carry out detailed modeling of
the discs.  Rather, we use a constant dust temperature, $T_{\rm d}=20$\,K,
which is the average found for Taurus discs by \citet{2005ApJ...631.1134A}
and assume a dust grain opacity,
$\kappa_\nu=10(\nu/1200\,{\rm GHz})\,{\rm cm}^2\,g\e$,
which follows \citet{1990AJ.....99..924B} but excludes the gas-to-dust ratio.
At the assumed distance, $d=420$\,pc, we therefore derive the following
approximate relation between SCUBA-2 flux density,
$F_{850}$ in mJy, and dust mass,
$$M_{\rm dust} = {F_\nu d^2\over\kappa_\nu B_\nu(T_{\rm d})}
  = 5.0\times 10^{-6}F_{\rm 850}\,M_\odot,$$
where $B_\nu$ is the Planck function.
We refer to a disc mass, $M_{\rm disc} = 100M_{\rm dust}$,
through the explicit inclusion of an ISM gas-to-dust ratio of 100.
The reason for factoring this out from $\kappa_\nu$ is that, ultimately,
we hope to be able to measure disc gas masses independently.

Under these assumptions,
the rms level of 2.9\,mJy\,beam\e\ within the central region of
the map translates to a $3\sigma$ detection limit of
$4.3\times 10^{-3}\,M_\odot \simeq 4.5\,M_{\rm Jup}$.
For the 9 detections, the inferred disc masses range from
$\sim 5-17\,M_{\rm Jup}$,
which lie at the lower end of the estimates for the minimum mass necessary
to form the solar system determined by \citet{1977MNRAS.180...57W}
and less than the more recent determination of $21\,M_{\rm Jup}$
by \citet{2005ApJ...627L.153D}.
The 83 YSOs with infrared excesses that were not individually
detected have an average (stacked) flux density of 1.0\,mJy\,beam,
which corresponds to a disc mass of $0.54\,M_{\rm Jup}$.
We therefore conclude that, despite their large infrared excesses,
Jupiter scale giant planet formation must be largely complete in
these objects.  For the 190 Class III YSOs, the non-detection
of the mean emission provides a $3\sigma$ limit to the disc mass
of less than $0.24\,M_{\rm Jup}$, or in terms of dust only,
$M_{\rm dust} < 0.75\,M_\oplus$.

As we only observed the 6 brightest SCUBA-2 discs in the cluster with the SMA,
we can draw only very limited conclusions about their gas content.
However, the fact that we detect CO in four of these discs demonstrates that
at least some molecular gas remains in these systems.
The CO emission is almost certainly optically thick so it is not
possible to estimate the gas mass from this line alone.
However, the existence of CO requires a minimum molecular hydrogen
column density, $N_{\rm H_2}>10^{21}$\,cm\ee\ \citep{2011ApJ...740...84Q},
which implies a lower limit to the gas mass,
$M_{\rm gas}\simgt 10^{-2}(R/100\,{\rm AU})^2\,M_{\rm Jup}$,
for a disc with radius $R$.
The lack of a $^{13}$CO detection, which we might expect to have
a lower opacity, at a $3\sigma$ sensitivity of
$\sim 0.5$\,Jy\,\kms\ translates to a limit on the average
column density,
$N_{\rm ^{13}CO}\simlt 10^{15}$\,cm\ee\ in the $2\farcs 5$ SMA beam,
for a range of excitation temperatures $20-100$\,K.
Assuming abundances [H$_2$]/[CO] = $2\times 10^4$ and
[CO]/[$^{13}$CO] = 70, consistent with recent measurements in star forming
regions by \citet{2013MNRAS.431.1296R}, implies an upper limit to the gas mass,
$M_{\rm gas}\simlt 0.5\,M_{\rm Jup}$.
This is an order of magnitude lower than the disc masses inferred from the
dust continuum and may indicate that the gas-to-dust ratio
in these discs is substantially less than the ISM value of 100 or that
most of the CO is frozen-out.  Actual detections of the CO isotopologues,
however, and detailed modeling are required to be definitive.

\subsection{Comparisons with other regions}
Although there were few detections, this submillimetre survey
of protoplanetary discs is the largest of its kind.
We only detect the peak of the mass distribution in \SO,
finding it to be about an order of magnitude lower
than that in the younger Trapezium Cluster in the same Orion
star forming complex \citep{2010ApJ...725..430M},
and the Taurus and $\rho$\,Ophiuchus star forming regions
\citep{2005ApJ...631.1134A, 2007ApJ...671.1800A}.

\begin{figure}
\centering
\includegraphics[width=80mm]{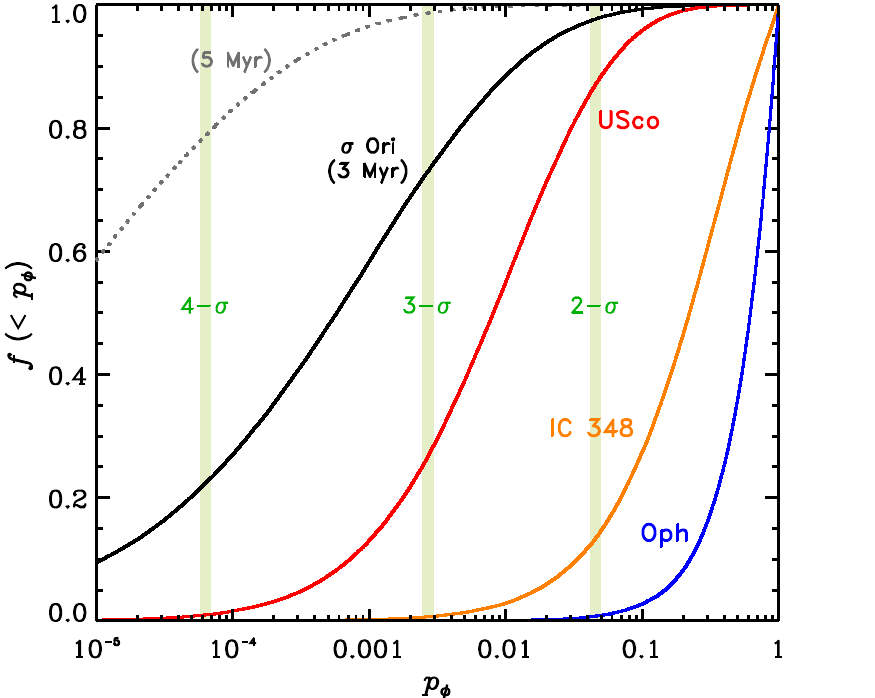}
\caption{The comparison of the disc submillimetre luminosity distributions
of \SO\ and Taurus.  $p_\phi$ is the probability that the two distributions
are drawn from the same parent population, taking into account the correlation
between disc mass and stellar spectral type, and allowing for upper limits.
The cumulative distribution of $p_\phi$ is calculated for $10^5$
Monte Carlo trials, the vast majority of which have a very low
probability of similarity.  The median $p_\phi$ is $5\times 10^{-4}$,
implying a $>3\sigma$ difference and we conclude that \SO\ disc masses
are significantly lower than those in Taurus.
The results for a similar analysis applied to the millimetre surveys
of discs in $\rho$\,Ophiuchus, IC\,348, and Upper Scorpius are from
\citet{2013ApJ...771..129A} and shown for reference.}
\label{fig:compare}
\end{figure}

Statistical comparisons between different regions are complicated
by the strong dependence of disc masses on their host stellar
properties, particularly mass and binarity \citep{2013ApJ...771..129A}.
Unfortunately, little information about either of these is
available for most members of \SO.
To compensate as best as we can, we use the 2MASS J band magnitudes
to estimate spectral types by comparing to
pre-main-sequence tracks \citep{2000A&A...358..593S}
at an assumed age of 3\,Myr and distance of 420\,pc.
To remove an additional model dependent step, we directly compare
$850\,\mu$m luminosities rather than inferred disc masses,
scaling Taurus measurements with the average millimetre SED slope
$F_\nu\propto\nu^{2.4}$ for that region \citep{2005ApJ...631.1134A}.
Most of the Taurus measurements used here are at $887\,\mu$m
so the uncertainty introduced by this extrapolation is minor.
Following the methodology in \citet{2013ApJ...771..129A},
we then sample from the Taurus disc luminosity distribution at each
spectral type to create $10^5$ synthetic distributions
compatible with the stellar properties of \SO.
These are then compared to the actual SCUBA-2 observations
using a standard two-sample test for censored datasets
\citep{1985ApJ...293..192F},
yielding a null probability that they are
not drawn from the same parent distribution, $p_\phi$.
The cumulative probability distribution, $f(<p_\phi)$, plotted in
Figure~\ref{fig:compare}, indicates that the \SO\ $850\,\mu$m disc
luminosity distribution is different from Taurus at $>3\sigma$ significance.

The significance decreases slightly if the \citet{1998A&A...337..403B}
pre-main-sequence tracks are used, since these predict lower
effective temperatures for a given J band absolute magnitude at 3\,Myr,
and the Monte Carlo simulations therefore sample from a lower set
of Taurus disc masses.
On the other hand,
the significance increases if the cluster is older since the inferred
effective temperatures are now higher and the synthetic disc flux densities
generally higher.  The dashed line in Figure~\ref{fig:compare} shows the
probability distribution if \SO\ is 5\,Myr old and 350\,pc distant.

The ages of star forming regions at early times, $\simlt 10$\,Myr,
are not well determined as
they depend on the uncertain initial conditions of protostellar birth models.
However, the general decline of the infrared disc fraction with time
is well characterised and can serve as a benchmark for the evolutionary
state of different regions.
Based on this simple observable, therefore, it is clear that \SO,
with a disc fraction of 27\%, is more evolved than
Taurus, with a disc fraction of 63\% \citep{2005ApJ...629..881H}.

Millimeter surveys toward other other moderately evolved
star forming regions, IC\,348 \citep{2011ApJ...736..135L}
and  Upper Scorpius \citep{2012ApJ...745...23M},
also found very low disc masses which were interpreted as
signatures of evolution.  At face value, the disc luminosity
distributions are similar to \SO\ (and very different from Taurus).
These surveys, and that of the younger $\rho$\,Ophiuchus from 
\citet{2007ApJ...671.1800A}
are included in Figure~\ref{fig:compare} following the same Monte Carlo
sampling comparison with the Taurus distribution.
$\rho$\,Ophiuchus and IC\,348 with disc fractions
$\sim 76$\% \citep{2009ApJS..181..321E} and
$\sim 47$\% \citep{2006AJ....131.1574L} respectively,
are statistically indistinguishable from Taurus.
The low observed disc masses in IC\,348 is
attributable to the generally low stellar masses in this region
\citep{1998ApJ...508..347L}.
Upper Scorpius, with a disc fraction of only 19\% \citep{2006ApJ...651L..49C},
is also not formally statistically different from Taurus.
In this case, because the intrinsic dispersion of disc masses at any
given stellar mass is a factor of 40 \citep{2013ApJ...771..129A},
the survey size of 37 sources is too small to rule out the
observed low disc masses being due to sampling.

The combination of sensitivity and large sample size of this SCUBA-2 survey,
covering a wide range of stellar masses, is essential to the statistical
significance of what might seem an intuitively obvious result:
as a young stellar region evolves and discs dissipate,
the dust mass of the remaining discs with infrared excesses decreases.  
Quantitatively, 
for our assumed distance and the standard disc dust opacity and gas-to-dust
ratio described above, the average Class II
disc mass decreases to about $0.5\,M_{\rm Jup}$ by about 3\,Myr.
Such low masses can only be reconciled with exoplanet statistics
if much of the (solid) material has been locked away in large bodies
with a relatively small surface area \citep{2010MNRAS.407.1981G}.

\subsection{Future directions}
With the exception of the Class III YSO, J0538-0225, which requires
independent verification, all the SCUBA-2 detected sources have
infrared excesses.  However, there is no particular signature
that correlates with the strength of the $850\,\mu$m emission.
Only 2 SCUBA-2 YSOs have IRAC excesses
that are greater than the median for Class II discs.
Furthermore, it is intriguing that the SEDs of 3 out of the 8
optically thick discs exhibit dips between $3.6\,\mu$m and $24\,\mu$m
(or more specifically a rising slope from $8\,\mu$m to $24\,\mu$m;
Figure~\ref{fig:sed}).
These transition disc signatures are generally quite rare,
comprising about 15\% of SED types in most clusters \citep{2011ARA&A..49...67W}.
Other examples of massive discs with depleted inner regions in several
Myr-old star forming regions
are J1604-2130 in Upper Scorpius \citep{2012ApJ...753...59M},
and the famous TW Hya \citep{2007ApJ...664..536H}.
The numbers are, as yet, too small to be statistically significant
but it is tempting to speculate that the reasons for the gap
and large outer disc reservoir are related.  One possibility
is a massive planet that produces enough torque on the outer
disc to slow down disc accretion \citep{2012ApJ...755....6Z}.

Looking ahead, the recently commissioned Atacama Large Millimeter Array (ALMA)
provides the capability to survey hundreds of discs in many different star
forming regions at an order of magnitude higher sensitivity in the line
and continuum than the data presented here.
The statistical positive signal of about 1\,mJy toward the
Class II non-detections is well within ALMA's sensitivities, even for
snapshot integrations. Unless the mass distribution is bimodal, therefore,
it should be able to detect most, if not all, Class II discs in \SO.
Moreover, it provides subarcsecond imaging and therefore the ability
to determine not only bulk masses but surface densities.
Ultimately such ALMA surveys of different regions will reveal the
evolution of the solid and gas components of discs as a function of
radius, stellar mass, and environment.
These will provide the key inputs required for population synthesis
models and understanding the diversity of exoplanetary systems.

\section{Summary}
\label{sec:summary}
We have mapped 297 (88\%) of the known members of the \SO\ cluster
at $850\,\mu$m.  Only 9 YSOs were detected with inferred
masses ranging from 5 to $17\,M_{\rm Jup}$.  The infrared SEDs of these
sources exhibit a wide diversity suggesting that the evolution of their
structural properties such as inner holes and flaring is
largely independent of their bulk dust mass.
Based on a stacking analysis, the mean mass of the
83 Class II YSOs that were not individually detected is $0.5\,M_{\rm Jup}$.
We followed up 6 sources with spectroscopic observations and detected CO
emission toward 4 of them, suggesting that the gas remains broadly coupled
with the dust as the discs dissipate.

The general weakness of the emission in the $\sim 3$\,Myr \SO\ discs
illustrates how rapidly the amount of raw material available for
planet formation decreases with time.
Whereas 27\% of the cluster members have optically thick discs,
the average mass of 90\% of these discs is only about half a Jupiter mass.
This suggests a yet faster timescale for giant planet formation than
the oft-used infrared lifetime plot.
SCUBA-2 has proven to be an efficient survey instrument for large
clusters to learn about the collective evolution of discs and as
a finder telescope for followup interferometry.
As ALMA begins millimetre wavelength surveys of the dust and gas content
of protoplanetary discs we can expect more precise constraints
on planet formation models.

\section*{Acknowledgments}
This work is supported by funding from the NSF through grant AST-1208911.
We thank Remo Tilanus for advice with the data reduction and the
referee, Jane Greaves, for her comments.
The James Clerk Maxwell Telescope is operated by the Joint Astronomy Centre
on behalf of the Science and Technology Facilities Council of the United
Kingdom, the Netherlands Organisation for Scientific Research, and the
National Research Council of Canada.  Additional funds for the construction
of SCUBA-2 were provided by the Canada Foundation for Innovation.
The Submillimeter Array
is a joint project between the Smithsonian Astrophysical Observatory and the
Academia Sinica Institute of Astronomy and Astrophysics and is funded by
the Smithsonian Institution and the Academia Sinica.
This research has made use of the SIMBAD database,
operated at CDS, Strasbourg, France and the 
Two Micron All Sky Survey, which is a joint project of the
University of Massachusetts and the Infrared Processing and Analysis
Center/California Institute of Technology, funded by NASA and NSF.


\bsp
\label{lastpage}

\end{document}